# Gravitational Wave for a pedestrian

Asis Kumar Chaudhuri
Variable Energy Cyclotron Centre
Kolkata-700 064

**Abstract:** The physics of gravitational wave and its detection in the recent experiment by the LIGO collaboration is discussed in simple terms for a general audience. The main article is devoid of any mathematics, but an appendix is included for inquisitive readers where essential mathematics for general theory of relativity and gravitational waves are given.

## *Introduction*

On February 11, 2016, LIGO (which is acronym for the Laser Interferometer Gravitational-Wave Observatory) collaboration announced the detection of Gravitational wave [1]. Following the announcement, there is considerable interest among the general populace about the discovery. They wanted to know more about the gravitational wave. What is it? Why its discovery is important? What are the implications? As a practitioner of science, this widespread enthusiasm to a scientific discovery is gratifying to me. Possibly, this is the second time in the World's history that the general public got interested in a purely scientific discovery. The earlier one was the discovery of bending of light. In 1915, Albert Einstein formulated a theory of gravity which is generally known as the General Theory of Relativity and predicted that light, which generally travel in a straight line, will bend in presence of a heavy mass. In 1919, during a solar eclipse, Sir Aurther Eddington and his collaborators measured the bending of light due to solar mass. Incidentally, Einstein also predicted the gravitational wave and present excitement is about the experimental verification of his prediction.

This article is an attempt to satiate common men's urge for knowledge about the gravitational wave. It is not for an expert. A proper understanding of gravitational wave and its detection mechanism requires some mathematical knowledge, which a general reader may be lacking. It will be my endeavor here to explain the gravitational wave and its detection, devoid of any mathematics. However, for an inquisitive reader, at the end, I have added an appendix where little bit mathematics of general relativity is given. I believe in my reader's intelligence and hope that he/she will be patient enough to go through the appendix and in the process gain more insight. More information on the subject may be obtained in [2,3,4].

## *What is Gravity?*

We all know about gravity. Gravity is the force that makes the Earth rotate about the Sun, it is also the force that makes an apple fall to the ground. From antiquity, men knew about gravity. In 1596 Johannes Kepler (27 Dec. 1571- 15 Nov. 1630) published his book, Mysterium Cosmographicum, or Cosmographic Mystery, where he tried to give a geometric description of the then known Universe. In Mysterium Cosmographicus, Kepler took the first tentative step toward the modern picture, where Sun—by its gravitation—controls the motions of the planets. In the first edition, Kepler attributed to the Sun a *motricem animam* ("moving soul"), which causes the motion of the planets. Aristotle used the word to indicate "soul" or life to the animates which enables a living thing to do what it properly does—a plant to grow, a dog to run and bark, a person to talk and think. In the same vein, Sun has a soul which controls the planets. In the second edition of Mysterium Cosmographicum Kepler came further close to the modern picture, he supposed that some force—which, like light, is "corporeal" but "immaterial" emanate from the Sun and drive the planets. The proper understanding of force of gravity is due to Sir Isaac Newton, the English physicist, mathematician, and philosopher. For years he contemplated on the force that keep the planets orbiting around the Sun. As the legend goes, one day, while sitting under an apple tree, an apple fall on his head and the solution strikes him like a lightning. Newton asked himself, why the apple fall? It falls because the apple is attracted by the earth. Apple is also attracting the Earth, but its effect on earth is imperceptible due to its huge mass. He contemplated similar force of attraction between the heavenly bodies. Newton then formulated his famous universal law of gravitation,

*Any two objects exert a gravitational force of attraction on each other. The magnitude of the force is proportional to the product of the gravitational masses of the objects, and inversely proportional to the square of the distance between them. The direction of the force is along the line joining the objects.*

The concept of gravity underwent radical change when the German Scientist, Albert Einstein, successively proposed two theories; the special theory of relativity and general theory of relativity. Special theory of relativity altered our concept of space, time, energy and mass. In Newtonian mechanics, they are separate entities, but no-longer in relativity. Mass and energy are equivalent. Possibly, the most well known equation in the world is Einstein's mass-energy relation, $E=Mc^2$, mass times the square of the speed of light (c) is energy. Special theory of relativity also altered our concept of space and time. In classical or Newtonian physics space and time has separate identity. Newton wrote;

"*Absolute space, in its own nature, without relation to anything external, remains always similar and immovable;*" and "*Absolute, true and mathematical time, of itself, and from its own nature, flows equably without relation to anything external.*"

But in relativity, space and time do not have separate identity, rather they are part of a single entity called space-time continuum, continuum because in our experience there is no void in space or in time. What is the meaning of a single entity space-time continuum? Say, two observers are observing an occurrence or event, e.g. blossoming of a flower. If the two observers had their clocks synchronized, then depending on their position, they may argue about the location of the occurrence (one may say that the flower was at his left side, the other may say that the flower was at his right side), but both will agree with the time flower blossomed. In Newtonian mechanics, flow of time is same for both the observers. However, in special theory of relativity, even with synchronized clocks, two observers will clock different time, depending on their position.

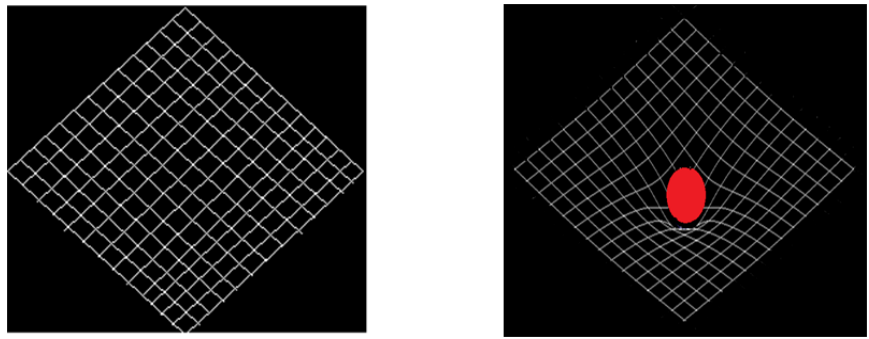

*Figure 1: A two dimensional view of flat and curved space time. Matter or energy curve the space-time in vicinity of it.*

Einstein's general theory of relativity also radically changed our understanding of gravity. According to Newton, gravity is the force of attraction between two massive objects. No so in Einstein's general relativity. In general relativity, gravity is related to the geometry of the space-time. A pictorial depiction of the space-time continuum, in Einstein's general relativity, in absence and in presence of mass is shown in Figure 1. In general relativity, in presence of mass, space-time is curved and gravity is nothing but the curvature of the space-time. For example, if asked, why the Moon does not fly off into the space, rather than orbiting around the Earth, Newton would have answered that the force of gravity acting between the Earth and Moon, hold it in the orbit. On the other hand, Einstein would have replied that the Earth's mass bends the space and time around itself and moon follows the curves created by the massive Earth.

In Einstein's theory, in a sense, the space-time itself is the gravity. Now, mathematically space-time is completely known if an associated quantity called "metric tensor" is known. Concept of metric tensor is rather complicated and I will not discuss it here, but in relativity, the metric tensor has 16-components out of which only 10 can be independent (meaning remaining 6 can be written in terms of those 10 components). The 10-component metric tensor codifies all the geometric information of the space-time. In physics, by Einstein's equation one generally means his general relativity equation. Einstein's equation is a relation between the metric tensor defining geometry of the space-time and mass-energy content of the Universe. The equation shows that the structure of the metric tensor changes in presence of mass or energy and that the change automatically involves Gravitational constant G. In absence of matter, space-time is flat and the metric tensor has one form. If matter is put into, the metric tensor changes to another form.

## *What is gravitational wave?*

We all have some idea of wave. When you throw a pebble on otherwise a still pond, a wave forms; starting from the origin (the point where the pebble touched the water) it spread across the pond. We call it wave as it has some pattern and the pattern repeats again and again. How a wave is formed is best understood if you had seen a Mexican Wave. Mexican wave was first seen in 1986 World Cup in Mexico (hence the name). In Mexican wave, people in a large stadium briefly stand up then sit down, each person doing so immediately after the person on one side of him. From a large distance, you see a wave swiftly moving across the stadium. Nobody moves from its position, they are simply standing up and sitting down. The same thing happens in a wave in the pond. Water particles are just going up and down like the people in the stadium, but the wave moves. Scientifically speaking, a wave is an oscillatory propagation of a disturbance, transferring energy from one point to another. There are several types of waves, e.g. transverse wave when the wave moves perpendicular to the direction of motion of the constituents as in the wave in a water, longitudinal wave, the wave moves parallel to the direction of motion of the constituents as in a sound wave. Mathematically, when the variation of disturbance with space and time obey certain relationship, the disturbance is said to be propagating as a wave. In our everyday experience, a wave requires a material medium. For example, in the propagation of sound wave, air is the material medium; layers of air columns alternately compressing and relaxing produces sound. In a wave on a pond, water is the material medium. There are waves that do not require any material medium. For example, light is a wave. Light is oscillation of electric and magnetic field and can propagate in a material medium as well as in an empty space. In all the cases discussed there is space in the back ground. The water body is occupying some space, the electric and magnetic fields exists in space.

In relativity, space-time is the gravity. Consider we are disturbing the space-time e.g. by rotating a heavy mass. As the disturbance of a material medium can travel as a wave, the disturbance of the space-time can also travel. If it travels then we get gravitational waves. Gravitational waves are ripples in the space-time fabric of the Universe. An artist's concept of gravitational wave is shown in *Figure 1*. In 1915, Einstein proposed his general theory of relativity and in 1916 he predicted existence of gravitational wave. Now, Einstein's general relativity equations are non-trivial equations and hard to solve under general conditions. It is a complex set of 10-non-linear equations. Einstein solved the equations in the weak field limit. What is the weak field limit? In presence of mass, space-time is curved, i.e. space-time has finite curvature. Space-time is flat in absence of mass and the curvature is zero. In the weak field limit, the space-time is very close to a flat space-time such that the curvature differs minimally from zero. In other words, metric tensor of the space-time departs minimally from the metric tensor of a flat space-time. The difference can be called perturbation. When Einstein solved his equation in the weak field limit, he obtained a mathematical equation called wave equation. The equation can be interpreted as propagation of the space-time perturbation as a wave with speed of light.

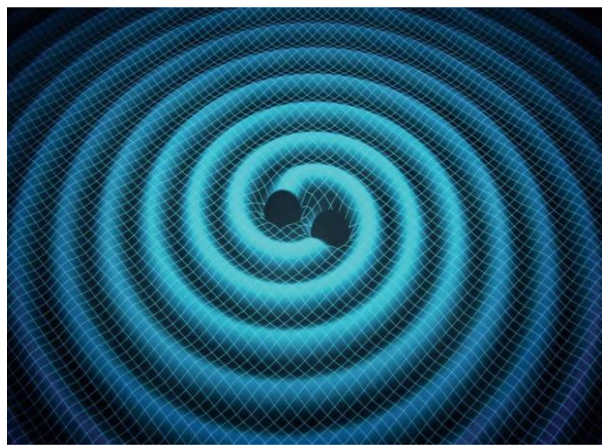

*Figure 2: Artist's depiction of gravitational wave from a binary neutron star. Image curtsey NASA.*

Even though Einstein predicted gravitational wave, initially, there was some confusion whether or not gravitational waves carries energy and is detectable. The confusion arises due to rather non-trivial character of Energy in general relativity. While in special relativity, energy is conserved, in general relativity, conservation of energy is not simple. Conservation of any quantity is the result of an underlying symmetry. For example, momentum is conserved if there is spatial translational symmetry, i.e. is nothing changes if the system under consideration is shifted by certain amount. Similarly, energy is conserved if the system in invariant under time. In general relativity, time is part of the coordinate system, and in general depends on the position. Then, globally, energy is not conserved. However, any curved space-time can be considered to be locally flat and locally, energy is conserved. Einstein himself was not sure about the reality of gravitational wave. In 1936, he and one of his young collaborator, Nathan Rosen solved the equations and arrived at a conclusion that the gravitational waves do not exist. They wrote a paper titled "Do gravitational wave exist?" and submitted to the leading American Journal, Physical Review. There is an interesting episode related to this paper. It was reviewed by a young scientist, Howard Percy Robertson. Robertson did not agree with Einstein-Rosen's conclusion and wrote a 10 page report showing the problems in the paper. Einstein was angry. To the Editor of Physical Review he wrote,

*"We (Mr. Rosen and I) had sent you our manuscript for publication and had not authorized you to show it to specialists before it is printed. I see no reason to address the — in any case erroneous — comments of your anonymous expert. On the basis of this incident I prefer to publish the paper elsewhere."*

Einstein vowed never again to submit a paper to the Physical Review and indeed, he kept his vow, never again submitted a paper to Physical Review. Later, Einstein reversed his opinion about gravitational wave and published a much altered version "On gravitational wave" in a lesser known journal, Journal of the Franklin Institute. Abstract of the paper is revealing, and I reproduce it for historical interest.

*"The rigorous solution for cylindrical gravitational waves is given. For the convenience of the reader the theory of gravitational waves and their production, already known in principle, is given in the first part of this paper. After encountering relationships which cast doubt on the existence of rigorous solutions for undulatory gravitational fields, we investigate rigorously the case of cylindrical gravitational waves. It turns out that rigorous solutions exist and that the problem reduces to the usual cylindrical waves in Euclidean space."*

Even though Einstein accepted reality of gravitational wave, Nathan Rosen, till his death, held to his belief; Gravity waves are unreal.

The confusion whether or not Gravitational wave is real continued for long. In 1957, in a Relativity meeting at Chapel Hill, Richard Feynman presented an argument in favor of the gravitational wave. The argument is now known as "sticky bead argument." It is a thought experiment. Imagine two beads attached on a rod. The beads are free to slide. It the stick is held transversely to the wave's direction of propagation, the wave will generate a tidal force about the midpoint of the stick. These produce alternating, longitudinal tensile and compressive stresses in the material of the stick; and the beads, being free to slide, move back and forth in response to the tidal forces. If contact between the bead and stick is 'sticky', then heating of both parts will occur due to friction. According to Feynman the heating indicates that the gravitational wave did indeed impart energy to the bead and rod system. So it must have indeed transported energy.

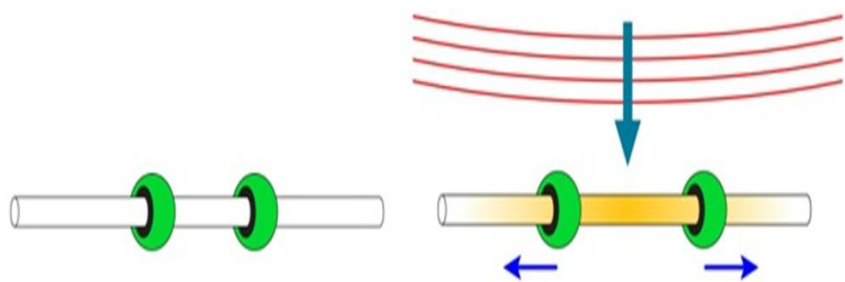

*Figure 3: Illustration of sticky bead argument. See the text for explanations.*

Even though Feynman's argument proved reality of gravitational waves, there was question of measurement. Gravitational wave changes the distances between two test masses. But how to measure that change? Say you are using a ruler to measure a given distance and you measure it to be 4 ruler lengths. Gravitational wave will change the given distance, but it will also change the ruler length. Then, you always measure the given distance 4-ruler lengths in presence or absence of gravitational wave. Much later it was realized that the change in length induced by the gravitational wave can be measured, if instead of the distance, we measure the time taken by light to traverse the distance. The speed of light is constant and is unaffected by the gravitational wave.

## *How a material body is affected by the gravitational wave?*

Gravitational waves are waves of perturbation of a flat space-time metric. The perturbation has 10-components, each of which can propagate as a wave. Under certain conditions of the 10-components only two survives. Effect of a gravitational wave of one of these two components is rather spectacular. It produces opposite effects on the two axes, contracting one while expanding the other. As shown in *Figure 4*, if a ring of particles lie on the plane of the paper, and the wave comes from the top, alternately, the ring will be squeezed and elongated. The wave changes the distance between two test particles. In experiment, one tries to measure the change. But what is the order of change expected in a given length due to gravitational wave?

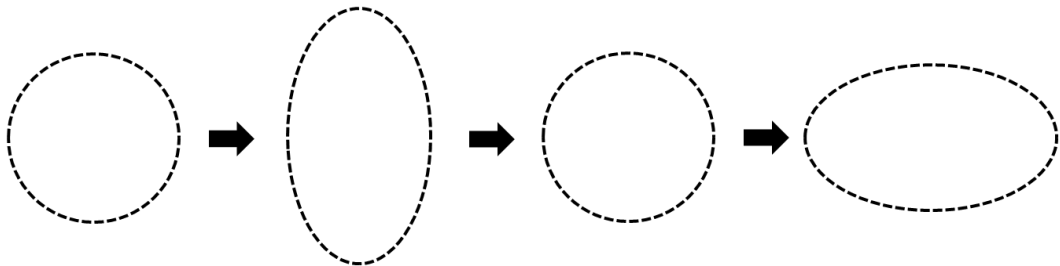

*Figure 4: Effect of gravitational waves on a ring of particles.*

Any accelerating mass sends off gravitational wave, however, unless the body is really massive it will not be detectable. Binary neutron stars or collision of two black holes etc. are typical source of a gravitational wave. Indeed, first indirect evidence of gravitational wave was obtained from binary neutron stars. Neutron stars are compact objects with mass of the order of 1-2 solar mass but with radius of the order of 10 Km. How compact it is can be understood if we note the solar radius is 695,500 Km. In binary neutron stars, two stars rotate about a common centre of mass. As they rotate, they send of gravitational wave and in the process lose energy and get closer and closer. It is called inspiralling. As they get closer they send off more gravitational waves and gets even closer, eventually colliding with each other. Just before the collision, the will send off intense gravitational wave. Scientists have estimated the strength of the gravitational wave send off by a binary neutron star system. A binary neutron star system, each with mass 1.4 times the solar mass, orbiting in a circular orbit of radius 20 Km, with orbital frequency 1000 Hz, at a distance R=15 Mpc[1] will produce a metric perturbation of the order of

---

[1] *Astronomical distances are measured in unit of parsec (symbol pc). 1 pc=3.09x10$^{16}$ meters. Mpc=3.09x10$^{22}$ meters.*

$$h \sim 6 \times 10^{-21}.$$

Relative change in length (called strain) is then extremely small $\Delta L/L \sim 10^{-21}$. Strain of gravitational wave from merger of two black holes is also calculated to be of the same order. It explains why physicists and astronomers regard the $10^{-21}$ threshold as so important. Instruments must be sensitive enough to detect this minute change in relative length.

## *Principle of detection of gravitational wave:*

Measuring a strain of the order of $10^{-21}$ means that a 1 cm length is required to be measured within an accuracy of $10^{-21}$ cm. How small it is can be understood if we compare it with the average size of an atom or nucleus. Average size of an atom is $10^{-8}$ cm, and average size of a nucleus is $10^{-15}$ cm.

LIGO collaboration detected the strain of the order of $10^{-21}$ in a Michelson interferometer. Michelson interferometers are built on a wave property called interference. If two waves superpose their amplitudes adds up. If crests fall upon crests and troughs fall upon trough, the amplitude increases and it is called constructive interference and if crests fall upon troughs and troughs fall upon crests, two waves destructs each other and we call it destructive interference. Pictorially, the phenomenon is shown in *Figure 5*.

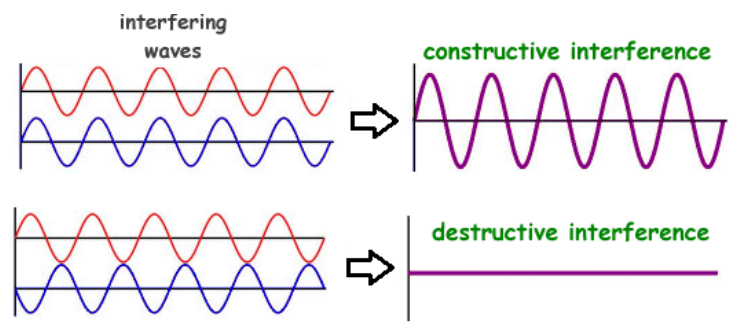

*Figure 5:    Constructive and destructive interference of two waves.*

Basic principle of the Michelson interferometer is explained in the *Figure 6*. Light of known wave length from a laser source fall on a beam splitter (a half polished mirror) which allow 50% the light to pass through and 50% reflected at 90 degree. The transmitted and reflected light beams travel through two arms of the interferometer to the mirrors $M_1$ and $M_2$ to get reflected back and merge at the splitter. The splitter then reflects the merged beam to a detector, say a photo diode. In merging, light waves of two beams superpose on each other. Depending on the path lengths travelled by the two beams, essentially, the distance of mirrors $M_1$ and $M_2$ from the beam splitter, interference patter will be formed at the detector. A small mismatch in the travel length will result in a fringe pattern (alternate rings or light and dark shades). The separation of the fringes depends on the wave length and path length difference of the two light beams. For a known wavelength light, the fringe separation gives the difference in length traversed by the two light beams.

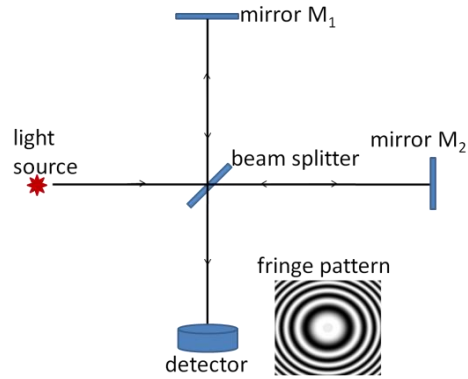

*Figure 6: Schematic diagram of a Michelson interferometer.*

With reference to *Figure 4*, effect of the gravitational wave on Michelson interferometer can be understood. It is illustrated in *Figure 7*. Let the two mirrors $M_1$ and $M_2$ are placed at a distance of L from the splitter. Let the gravitational wave propagate along the Z-axis. As shown in the *Figure 7*, the gravitational wave will set the mirrors to oscillation, the lengths changing from L to L+ΔL and L-ΔL. The change of length ΔL will result in an interference pattern and the induced change ΔL can be measured.

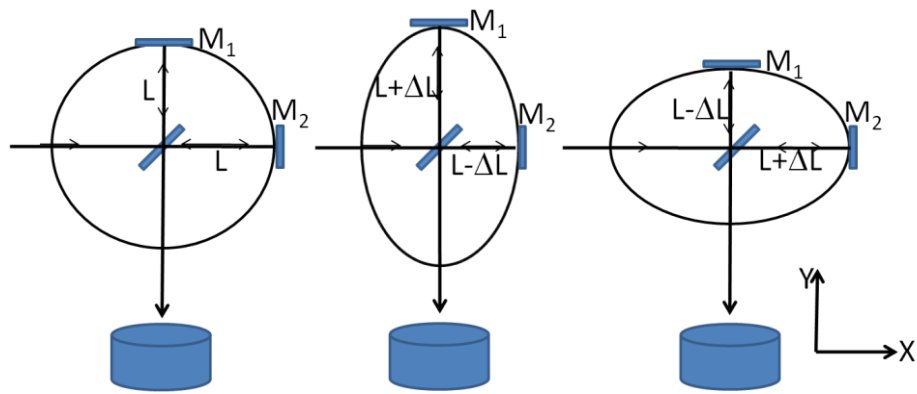

*Figure 7: Cartoon illustration of effect of gravitational wave on two mirrors of a Michelson interferometer.*

But isn't laser is also affected by gravitational wave as much as the arm lengths? Yes, they are. But, nowhere in Michelson Interferometer is laser used to measure a distance. We measure the time taken by the laser to travel between the arm lengths. That time is unaffected. Speed of light is constant and a blue light and a red light will traverse a given distance is the same time.

## *LIGO: an experimental marvel*

LIGO is a huge experiment where approximately 1000 scientists from all over the World collaborated to set up a giant experiment, approximately costing 1 billion US dollars. In late 1960's and early 1970's, a group of scientists at Massachusetts Institute of Technology (MIT) and California Institute of Technology (Caltech) started discussions about setting up an experiment to detect gravitational wave. In 1980, the US National Science Foundation (NSF) funded the construction of two prototype interferometers; one at Caltech and one at MIT. It also asked for technical and cost study for a several-kilometer-long interferometer. With this study demonstrating the feasibility of long interferometers and prototypes interferometers showing success, in 1984 Caltech and MIT joined together for the joint design and construction of LIGO and in 1990 NSF funded the project. The collaboration decided to built two identical Michelson interferometer in two  locations; one at Hanford, Washington and the other at Livingston, Louisiana  separated by 3,002 km. It was absolutely essential to built two detectors at widely separated locations. Essentially, LIGO measures the path length difference of light between two arms of the interferometer. A local vibration due a small earth quake can induce sufficient perturbation to change the path length. With two widely separated interferometers the accidental perturbations due to seismic vibration can be eliminated.  Initial LIGO detectors, operated between 2002 and 2010 did not detect any gravitational waves. However, it provided for the crucial understandings for augmentation of the detection capabilities to install advanced LIGO.

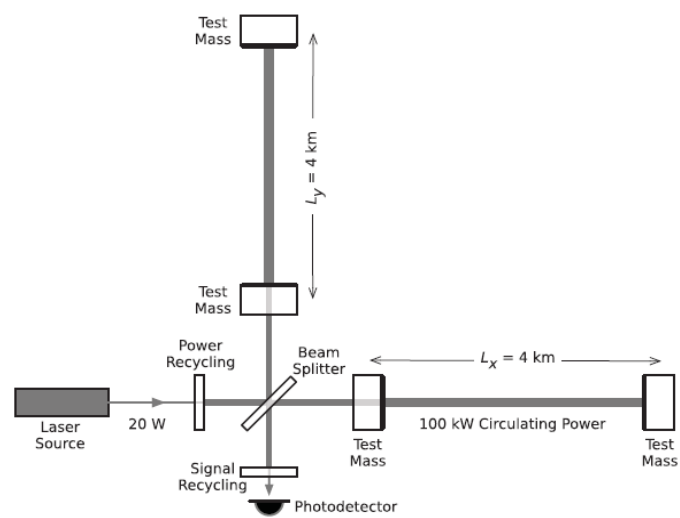

*Figure 8: A schematic of Michelson interferometer used in LIGO experiment.*

A schematic of a Michelson interferometer in advanced LIGO is shown in *Figure 8*. LIGO called the mirrors test masses. To facilitate the small strain measurement, the mirrors are placed at a distance of 4 km from the beam splitter. Even 4-Km distance is not sufficient. LIGO collaboration advanced the detection capability of a Michelson interferometer by many folds by modifying the design of the interferometer to include "Fabry-Perot" cavities; using additional mirrors near the beam splitter. In Fabry-Perot cavities the laser beam oscillate back and forth between two mirror to effectively increase the path length to 1120 Km before merging two beams. In a sense, in LIGO's Michelson interferometer, the mirrors are placed at a distance of 1120 Km. LIGO's mirrors are aligned such that in the absence of gravitational wave, no light will reach the detector.

The quest that began in late 1960, ended on September 14, 2015 at 9.50 UTC (coordinated Universal time, previously known as GMT). The twin detectors at Livingstone, Louisiana and at Hanford, Washington, observed a transient gravitational-wave signal. The signal sweeps upwards in frequency from 35 to 250 Hz with a peak gravitational-wave strain of $1.0 \times 10^{-21}$. It matches the waveform predicted by general relativity for the inspiral and merger of a pair of black holes of mass 29 solar mass and 36 solar mass at a distance of 410 Mpc. This is the first direct detection of gravitational waves and the first observation of a binary black hole merger.

Truely speaking, detection of gravitational wave was more of a technological challenge than that of physics challenge. The underlying physics was known for 100 years, method of detection was also known. Challenge was the unprecedented scale of precision ($10^{-21}$) required for the measurement. In a sense, LIGO experiment is an example of modern technological marvel. Some technological accomplishments of LIGO are mentioned below.

**Seismic isolation:**

LIGO, designed to detect $10^{-21}$ order of strain is extremely sensitive to all kinds of vibrations e.g. nearby movement of trucks, earthquakes. Extraordinary measures were taken to isolate the mirrors from this kind of spurious vibrations. Sensors, which can detect smallest vibration on earth, were attached to the mirrors. By a feedback system, counter movements were induced to the mirrors to keep them fixed in their positions. Additionally, the mirrors were attached to a 4-stage pendulum, which absorbs any movements not cancelled by the above method.

**Ultra-high vacuum system:**

The experiment required to create and sustain one of the purest vacuum systems on Earth. From the beam splitter, the two mirrors are placed at a distance of 4 km. This required construction of L-shaped vacuum chamber with volume of 10000 cubic meters. LIGO's vacuum volume is surpassed only by the Large Hadron Collider at CERN, Switzerland which has a vacuum volume of 15000 cubic meters. 1100 hours of pumping is required to obtain desired vacuum $10^{-9}$ torr[2] in the L-shaped chamber. The high quality vacuum is absolutely essential for the super-precision experiment. It prevent propagation of sound wave (sound cannot travel through vacuum) and disturb the mirrors. It also prevent any temperature variation inside the chamber and affect the quality of laser beams.

**LIGO mirrors:**

Each weighing 40 Kg, LIGO mirrors are special. Made of fused silica, each was polished to nano meter smoothness (i.e. in atomic level). Each laser in each arm travels about 1120 km before being merged with its partner. Perfect polishing is required to maintain the quality and stability of the laser beams. LIGO's main mirrors are also best ever constructed. It absorbs only one out of every 3.3 million photons, rest are reflected or transmitted.

**LIGO Laser:**

Laser used by LIGO is also special. Initial 4 watt, 808 nanometer laser from Nd-YAG (neodymium-doped yttrium aluminum garnet crystal) is amplified several times to produce a 700 watt 1064 nanometer wave length laser beam that enters into the interferometer. Inside the interferometer, its power is further increased to 100 kilo watt to circulate between the mirrors. LIGO laser is extra-ordinarily stable. By using feed-back mechanism the laser is made 100-million more stable than its intrinsic stability.

**Precision concrete pouring and leveling:**

Long arm lengths of the interferometer posed a problem; earth's curvature needs to be taken into account. Over a distance of 4-km, Earth is curved by 1 meter. The curvature was taken into account by most precise concrete pouring and leveling imaginable. Unless taken into account, laser light traveling 4-Km would have missed the mirrors.

---

[2] *Torr is a unit of pressure. 760 torr corresponds to standard atmospheric pressure.*

**High speed computer and data collection:**

LIGO produced an enormous amount of data. In a day's operation it generates terabytes (1000 gigabytes) of data, that required to be stored and process. Storing and processing of the vast amount of data require huge resource. LIGO harness a computer power that is equivalent to running a modern 4-core laptop computer for 1,000 years!

# *Conclusion:*

LIGO the most sensitive detector ever built detected gravitational wave. What next? At present LIGO consists of two identical detectors. It is planned to install another detector in India. Indian government formally agreed to bear the cost of the third detector. With three detectors in unison, LIGO will be able to identify the location of the source of gravitational wave with more precision.

What are the implications of the detection? The very first experiment gave a robust proof of existence of black hole and black hole merger. What lies ahead is uncertain. To the mankind, detection of gravitational wave has opened a window to look into Universe hitherto unknown. I will end this article with a quote by Dr. Kip Thorne, a renowned American physicist and one of the founders of LIGO:

*"With this discovery, we humans are embarking on a marvelous new quest: the quest to explore the warped side of the universe – objects and phenomena that are made from warped space-time. Colliding black holes and gravitational waves are our first beautiful examples."*

**Appendix:**

For an inquisitive reader, I have included little details of Einstein's theory. Einstein's general relativity equation is a tensor equation in a metric space. Understanding of tensor will be facilitated if we understand terms like scalar and vector. A scalar is a quantity that has only a magnitude or value. A vector is a quantity that has a value as well a direction. Our ordinary world is 3-dimensional. With three perpendicular reference directions (X,Y,Z), any arbitrary vector can be decomposed into three components. The components can be arranged in a one dimensional array. For example, three components; $A_x, A_y, A_z$, of a vector can be arranged in a row or column;

$$(A_x, A_y, A_z) \quad \text{or} \quad \begin{pmatrix} A_x \\ A_y \\ A_z \end{pmatrix}$$

A tensor is a quantity that is more general than a vector. Like a vector, it is also a multi-component quantity, but unlike a vector where the components can be designated by one index and a vector can be arranged in a one dimensional array, a tensor quantity requires more than one index to designate and can be arranged in two or more dimensional array. For example, a tensor requiring only two indices can be arranged in a 2x2 array of rows and columns,

$$\begin{pmatrix} A^{xx} & A^{xy} & A^{xz} \\ A^{yx} & A^{yy} & A^{yz} \\ A^{zx} & A^{zy} & A^{zz} \end{pmatrix}$$

Number of indices required to designate a tensor is called its rank. The tensor written above, requiring two indices is a rank 2 tensor. One can say; a vector is a rank 1 tensor and scalar is a rank 0 tensor.

Now we understand what is meant by tensor quantities, let us try to understand a metric space. A space with a notion of distance is called a metric space. Our ordinary world is a three dimensional metric space. Square of the distance (ds) between two adjacent points located at (x,y,z) and (x+dx,y+dy,z+dy) can be calculated as,

$$ds^2 = [(x+dx) - x]^2 + [(y+dy) - y]^2 + [(z+dz) - z]^2$$
$$= dx^2 + dy^2 + dz^2$$

In relativity space and time do not have separate existence; they are part of the space-time continuum. Relativistic world is 4-dimensional. In addition to three spatial directions, one temporal direction is required to describe a vector in 4-dimensional relativistic world. For mathematical reason, the temporal direction is taken to be complex; ict, i=√-1 and c=velocity of light. A general relativistic vector will require 4-components ($A_{ict}, A_x, A_y, A_z$) where $A_{ict}$ is the component value in the temporal direction and $A_x$, $A_y$, $A_z$ are the three spatial components.

Scientists have come up with an economical way to express vectors and tensors. Using the following identification;

$$x^0 \equiv ict;\ x^1 \equiv x;\ x^2 \equiv y;\ x^3 \equiv z,$$

a generic 4-vector can be written as $A^\mu$ or $A_\mu$, where $\mu$ can have any of the 4-values; 0,1,2,3 or equivalently, ict,x,y,z. In this notation, $A^0$ is the temporal and $A^1, A^2, A^3$ are spatial components of the vector. Similarly, a tensor can be written as $A^{\mu\nu}$ or $A_{\mu\nu}$ where $\mu$ and $\nu$ both can have any of the 4-values; 0,1,2,3 or equivalently, ict,x,y,z.

The notion of distance in three dimensions can be generalized to the 4-dimensional relativistic world. The distance between two adjacent points located at $(x^0,x^1,x^2,x^3)$ and $((x^0+dx^0),x^1+dx^1,x^2+dx^2,x^3+dx^3)$ can be written as,

$$ds^2 = -dx_{ict}^2 + dx_x^2 + dx_y^2 + dx_z^2$$
$$= -(dx^0)^2 + (dx^1)^2 + (dx^2)^2 + (dx^3)^2$$

The minus sign in front of temporal component arises due to the use of complex time coordinate. Using a 4x4=16 component tensor;

$$\eta_{\mu\nu} = \eta^{\mu\nu} = \begin{pmatrix} -1 & 0 & 0 & 0 \\ 0 & 1 & 0 & 0 \\ 0 & 0 & 1 & 0 \\ 0 & 0 & 0 & 1 \end{pmatrix},$$

you can do a little calculation to show that the distance can be written in a compact form,

$$ds^2 = \eta_{\mu\nu} dx^\mu dx^\nu$$

with the understanding that whenever a symbol is repeated, they have to be summed. A space where distance can be written in the above form is called metric space and the tensor $\eta_{\mu\nu}$ is call metric tensor. Metric tensor is also symmetric, i.e. interchanging the indices does not change the tensor, $\eta_{\mu\nu} = \eta_{\nu\mu}$. Mathematically, a metric tensor completely defines the structure of the space-time. Metric tensor $\eta^{\mu\nu}$ with the above structure, i.e. with diagonal components $\eta_{00}=-1$, $\eta_{11}=1$, $\eta_{22}=1$, $\eta_{33}=1$, and off diagonal components zero corresponds to a flat space-time. Metric $g^{\mu\nu}=g_{\mu\nu}$ for a general space-time can have non-zero values for all the components and, additionally they can be function of the coordinates (ict,x,y,z).

In general relativity, gravity is nothing but the curvature of the space-time. Curvature is a geometrical concept; it quantifies how much a surface deviates from a flat surface. Metric tensor $g^{\mu\nu}$ uniquely define space-time continuum and also its curvature. From the metric tensor one can derive a symmetric tensor called Ricci tensor $R_{\mu\nu}$. Trace R (or sum of the diagonal elements) of the Ricci tensor gives the curvature of the space-time continuum. Using the Ricci tensor, one can obtain Einstein tensor;

$$G^{\mu\nu} = R^{\mu\nu} - g^{\mu\nu} R$$

Einstein tensor has the property that

$$\frac{\partial}{\partial x^{\mu}} G^{\mu\nu} = 0$$

In our world energy and momentum are conserved. The law of conservation of energy-momentum can be written as,

$$\frac{\partial}{\partial x^{\mu}} T^{\mu\nu} = 0$$

where $T^{\mu\nu}$ is called the energy-momentum tensor and its different components are related to energy and momentum flow. For example, $T^{00}$ component corresponds to energy density, $T^{11}$ component corresponds to pressure or momentum flow etc. Identifying,

$$\frac{\partial}{\partial x^{\mu}} T^{\mu\nu} = \frac{\partial}{\partial x^{\mu}} G^{\mu\nu}$$

Einstein took the solution of these equations to be of the form,

$$G^{\mu\nu} = \kappa T^{\mu\nu}$$

the constant $\kappa$ can be found by demanding that the Newtonian gravity is recovered in the limit of a weak gravitational field and non-relativistic motion. One obtains, $\kappa = 8\pi G/c^4$ and we obtain the well known Einstein's equation,

$$R^{\mu\nu} - \frac{1}{2}g^{\mu\nu}R = \frac{8\pi G}{c^4}T^{\mu\nu}$$

Ricci tensor is identically zero in flat space-time and the equation is trivially satisfied in empty space, without any mass or energy. In presence of mass, $T^{\mu\nu}$ is non-zero, and the equation demands non-zero Ricci tensor, which effectively means that the space-time is not flat, it is curved.

Einstein obtained the gravitational wave equation in the weak field limit, when metric tensor can be written as,

$$g^{\mu\nu} = \eta^{\mu\nu} + h^{\mu\nu}, h^{\mu\nu} \ll 1.$$

$\eta^{\mu\nu}$=diag(-1,1,1,1) being the metric for the flat space-time, and $h^{\mu\nu}$ is the perturbation or the small departure from the flatness and depends on the space-time components, $h^{\mu\nu}=h^{\mu\nu}(t,x,y,z)$. When the metric is put into Einstein's equations, after certain manipulations, one obtains an equation for $h^{\mu\nu}$ which is exactly a wave equation in three dimensions;

$$\frac{\partial^2 h^{\mu\nu}}{\partial t^2} = c^2 \left( \frac{\partial^2 h^{\mu\nu}}{\partial x^2} + \frac{\partial^2 h^{\mu\nu}}{\partial y^2} + \frac{\partial^2 h^{\mu\nu}}{\partial z^2} \right)$$

Meaning of the above equation is simple. If the flat space-time is perturbed, the perturbation propagates like a wave with velocity of light c. Gravitational wave is oscillatory propagation of space-time perturbation $h^{\mu\nu}$. Under certain conditions of the 10-components of $h^{\mu\nu}$, only two survives and the perturbed metric is simplified as,

$$h^{\mu\nu} = \begin{pmatrix} 0 & 0 & 0 & 0 \\ 0 & h^{11} & h^{12} & 0 \\ 0 & h^{12} & -h^{11} & 0 \\ 0 & 0 & 0 & 0 \end{pmatrix}$$

Consider the distance between two test particles under the influence of the gravitational wave only of $h^{11}$ (when $h^{12}=0$). The distance can be calculated as,

$$ds^2 = (\eta^{\mu\nu} + h^{\mu\nu})dx_\mu dx_\nu$$
$$= -dt^2 + (1+h^{11})dx^2 + (1-h^{11})dy^2 + dz^2$$

Gravitational wave of $h^{11}$ produces opposite effects on the two axes, contracting one while expanding the other.

## *References:*

1. Observation of Gravitational Waves from a Binary Black Hole Merger, B. P. Abbott et al., Phys. Rev. Lett. 116 (2016) 061102.
2. www.logo.caltech.edu
3. An Introduction to General Relativity, Gravitational Waves and Detection Principles, Martin Hendry, Second VESF School on Gravitational Waves, Cascina, Italy, 2007.
4. Gravitational Waves, Sources, and Detectors, B. F. Schutz and F. Ricci, arXiv:1105.4735.